\RequirePackage{amsmath}

\documentclass[longauth]{aa}

\usepackage{verbatim}
\usepackage{graphicx}
\usepackage{txfonts}
\usepackage{subfigure}
\usepackage{float}
\usepackage{natbib}
\usepackage{twoopt}
\usepackage{xspace}
\usepackage{balance}
\usepackage{upgreek}
\usepackage{multirow}
\usepackage{mathrsfs}
\usepackage{etoolbox}

\makeatletter
\patchcmd\@combinedblfloats{\box\@outputbox}{\unvbox\@outputbox}{}{\errmessage{\noexpand patch failed}}
\makeatother

\usepackage[colorlinks=true, citecolor=blue]{hyperref}

\def\apjl{Astrophys. J. Lett.}
\def\apjs{Astrophys. J. Suppl.}

\def\aap{Astron. Astrophys.}

\newcommand{\Mpch}{$h^{-1}\,\mbox{Mpc}$\,}

\newcommand{\Om}{$\Omega_{\rm M}$\,}

\bibpunct{(}{)}{;}{a}{}{,} 
\DeclareGraphicsExtensions{.eps,.ps,.pdf}

\defcitealias{pierre2016}{XXL Paper I}
\defcitealias{pacaud2016}{XXL Paper II}
\defcitealias{lieu2016}{XXL Paper IV}
\defcitealias{eckert2016}{XXL Paper XIII}
\defcitealias{adami2018}{XXL Paper XX}
\defcitealias{pacaud2018}{XXL Paper XXV}
\defcitealias{chiappetti2018}{XXL Paper XXVII}
\defcitealias{plionis2018}{XXL Paper XXXII}

\begin{document}

\title{The XXL Survey \thanks{Based on observations obtained with
    XMM-Newton, an ESA science mission with instruments and
    contributions directly funded by ESA Member States and NASA. Based
    on observations made with ESO Telescopes at the La Silla and
    Paranal Observatories under programmes ID 191.A-0268 and
    60.A-9302.}  }

\subtitle{XVI. The clustering of X-ray selected galaxy clusters at
  $z\sim0.3$}

\author{
  F. Marulli\inst{\ref{1},\ref{2},\ref{3}}
  \and A. Veropalumbo\inst{\ref{4}}
  \and M. Sereno\inst{\ref{1},\ref{2}}
  \and L. Moscardini\inst{\ref{1},\ref{2},\ref{3}}
  \and F. Pacaud\inst{\ref{5}}
  \and M. Pierre\inst{\ref{6},\ref{7}}
  \and M. Plionis\inst{\ref{8},\ref{9}}
  \and A. Cappi\inst{\ref{2},\ref{10}}
  \and C. Adami\inst{\ref{11}}
  \and S. Alis\inst{\ref{12}}
  \and B. Altieri\inst{\ref{13}}
  \and M. Birkinshaw\inst{\ref{14}}
  \and S. Ettori\inst{\ref{2},\ref{3}}
  \and L. Faccioli\inst{\ref{6},\ref{7}}
  \and F. Gastaldello\inst{\ref{15}}
  \and E. Koulouridis\inst{\ref{6},\ref{7}}
  \and C. Lidman\inst{\ref{16},\ref{17}}
  \and J.-P. Le F{\`e}vre\inst{\ref{18}}
  \and S. Maurogordato\inst{\ref{10}}
  \and B. Poggianti\inst{\ref{19}}
  \and E. Pompei\inst{\ref{20}}
  \and T. Sadibekova\inst{\ref{6},\ref{7},\ref{21}}
  \and I. Valtchanov\inst{\ref{13}}
}

\offprints{F. Marulli \\ \email{federico.marulli3@unibo.it}}

\institute{
  Dipartimento di Fisica e Astronomia - Alma Mater Studiorum
  Universit\`{a} di Bologna, via Piero Gobetti 93/2, I-40129 Bologna,
  Italy\label{1}
  \and INAF - Osservatorio di Astrofisica e Scienza dello Spazio di
  Bologna, via Piero Gobetti 93/3, I-40129 Bologna, Italy\label{2}
  \and INFN - Sezione di Bologna, viale Berti Pichat 6/2, I-40127
  Bologna, Italy\label{3}
  \and Dipartimento di Fisica, Universit\`{a} degli Studi Roma Tre, via
  della Vasca Navale 84, I-00146 Rome, Italy\label{4}
  \and Argelander Institut f${\rm\ddot{u}}$r Astronomie,
  Universit${\rm\ddot{a}}$t Bonn, Auf dem H${\rm\ddot{u}}$gel 71,
  DE-53121 Bonn, Germany\label{5}
  \and IRFU, CEA, Universit{\'e} Paris-Saclay, F-91191 Gif-sur-Yvette,
  France\label{6}
  \and Universit{\'e} Paris Diderot, AIM, Sorbonne Paris Cit{\'e},
  CEA, CNRS, F-91191 Gif-sur-Yvette, France\label{7}
  \and National Observatory of Athens, Lofos Nymphon, Thession, Athens
  11810, Greece\label{8}
  \and Physics Department, Aristotle University of Thessaloniki,
  Thessaloniki 54124, Greece\label{9}
  \and Laboratoire Lagrange, UMR 7293, Universit{\'e} de Nice Sophia
  Antipolis, CNRS, Observatoire de la C{\^o}te d'Azur, 06304 Nice, France\label{10}
  \and Aix-Marseille Universit{\'e}, CNRS, LAM (Laboratoire
  d'Astrophysique de Marseille) UMR 7326, 13388, Marseille,
  France\label{11}
  \and Department of Astronomy and Space Sciences, Faculty of Science,
  Istanbul University, 34119 Istanbul, Turkey\label{12}
  \and Telespazio Vega UK for ESA, European Space Astronomy Centre,
  Operations Department, 28691 Villanueva de la Ca\~nada,
  Spain\label{13}
  \and H.H. Wills Physics Laboratory, University of Bristol, Tyndall
  Avenue, Bristol BS8 1TL, UK\label{14}
  \and Istituto di Astrofisica Spaziale e Fisica Cosmica di Milano, via
  Bassini 15, I-20133 Milan, Italy\label{15}
  \and Australian Astronomical Observatory, North Ryde, NSW 2113,
  Australia\label{16}
  \and The Research School of Astronomy and Astrophysics, Australian
  National University, Canberra, ACT 2611, Australia \label{17}
  \and Service d'{\'E}lectronique des D{\'e}tecteurs et d'Informatique,
  CEA/DSM/IRFU/SEDI, CEA Saclay, 91191 Gif-sur-Yvette, France\label{18}
  \and Osservatorio Astronomico di Padova, INAF, I-35141 Padova,
  Italy\label{19}
  \and European Southern Observatory, Alonso de C{\'o}rdova 3107,
  Vitacura, 19001 Casilla, Santiago de Chile, Chile\label{20}
  \and Ulugh Beg Astronomical Institute of the Uzbek Academy of
  Sciences, 33 Astronomicheskaya str., Tashkent, UZ-100052,
  Uzbekistan\label{21}
}

\date{Received --; accepted --}

\abstract 
{ Galaxy clusters trace the highest density peaks in the large-scale
  structure of the Universe. Their clustering provides a powerful
  probe that can be exploited in combination with cluster mass
  measurements to strengthen the cosmological constraints provided by
  cluster number counts. }
{ We investigate the spatial properties of a homogeneous sample of
  X-ray selected galaxy clusters from the XXL survey, the largest
  programme carried out by the {{\em XMM-Newton}} satellite. The
  measurements are compared to $\Lambda$-cold dark matter predictions,
  and used in combination with self-calibrated mass scaling relations
  to constrain the effective bias of the sample, $b_{eff}$, and the
  matter density contrast, $\Omega_{\rm M}$.}
{ We measured the angle-averaged two-point correlation function of the
  XXL cluster sample. The analysed catalogue consists of $182$ X-ray
  selected clusters from the XXL second data release, with median
  redshift $\langle z \rangle=0.317$ and median mass $\langle
  M_{500} \rangle\simeq1.3\cdot10^{14} M_\odot$. A Markov chain Monte
  Carlo analysis is performed to extract cosmological constraints
  using a likelihood function constructed to be independent of the
  cluster selection function.}
{ Modelling the redshift-space clustering in the scale range
  $10<r\,[$\Mpch$]<40$, we obtain $\Omega_{\rm M}=0.27_{-0.04}^{+0.06}$
  and $b_{eff}=2.73_{-0.20}^{+0.18}$. This is the first time the
  two-point correlation function of an X-ray selected cluster
  catalogue at such relatively high redshifts and low masses has been
  measured. The XXL cluster clustering  appears fully consistent with
  standard cosmological predictions. The analysis presented in this
  work demonstrates the feasibility of a cosmological exploitation of
  the XXL cluster clustering, paving the way for a combined analysis
  of XXL cluster number counts and clustering. } {}


\authorrunning{F. Marulli and the XXL Team}
\titlerunning{The XXL cluster clustering}

\maketitle


\section {Introduction}
\label{sec:introduction}

Galaxy clusters, the largest virialised structures in the present-day
Universe, provide one of the most powerful probes for constraining
cosmology. Their comoving number density is sensitive to both the
background geometry of the Universe and the growth rate of cosmic
structures \citep{allen2011}.  On the other hand, it is much harder to
exploit the clustering properties of galaxy clusters, due to the
challenging task of collecting large homogeneous cluster samples,
especially when the selection is done in the X-ray band
\citep{lahav1989, nichol1994, romer1994}.

Large samples of cosmic tracers are required to accurately describe
the underlying density field of the Universe. Wide galaxy surveys,
probing increasingly large and dense volumes of the Universe, have
played the primary role in this field  \citep[see
  e.g.][]{york2000, kaiser2010, dejong2013, guzzo2014, aihara2017}.

Despite the scarcity of cluster catalogues relative to galaxies, and
the difficulty in building up complete and pure samples covering wide
ranges of masses and redshifts, there are numerous advantages to
exploiting clusters as cosmic tracers.  Massive dark matter haloes
trace the rare highest peaks of the cosmological density field
\citep{kaiser1987}. Galaxy clusters, hosted by the most massive
virialised haloes, are more clustered than galaxies, with a clustering
signal that is progressively stronger for richer systems
\citep{klypin1983, bahcall1983, mo1996, moscardini2000, colberg2000,
  suto2000, sheth_mo_tormen2001}.  The capability of measuring
accurate cluster masses is crucial in order to constrain their
effective bias as a function of the cosmological model, something that
is not possible with galaxies and other cosmic tracers. Moreover,
clusters are relatively unaffected by non-linear dynamics at small
scales, so that the feature known as the Fingers of God in cluster
clustering is almost absent \citep{marulli2017}. The redshift-space
distortions at large scales also have a minor impact on cluster
clustering compared to galaxies due to their larger bias
\citep{kaiser1987, hamilton1992}. This simplifies the modelling of
cluster clustering, minimising the theoretical uncertainties in the
description of non-linear dynamics and redshift-space
distortions. Furthermore, the non-linear damping in baryon acoustic
oscillations of cluster clustering is small, thus improving the
significance of peak detection \citep{veropalumbo2014}.

Robust cosmological constraints have been obtained from the two-point
correlation function (2PCF) and power spectrum of optical and X-ray
selected galaxy clusters \citep[see e.g.][and references
  therein]{retzlaff1998, abadi1998, borgani1999, moscardini2000b,
  collins2000, schuecker2001, miller2001, balaguera2011, emami2017},
and even from baryon acoustic oscillations at large scales
\citep{estrada2009, hutsi2010, hong2012, hong2016, veropalumbo2014,
  veropalumbo2016}. The clustering of galaxy clusters has also been
analysed in combination with cluster number counts
\citep{schuecker2003, majumdar2004, mana2013} and gravitational
lensing measurements \citep{sereno2015} to strengthen cosmological
constraints and to break degeneracies.

The goal of this paper is to investigate the spatial properties of a
homogeneous sample of X-ray galaxy groups and clusters.  X-ray
selected cluster samples are less contaminated by projection effects
than optically selected ones, and can ensure a high level of purity.
This is crucial, in particular for cosmological investigations.  The
XXL survey, the largest programme carried out by the {\em XMM-Newton}
satellite to date, has been specifically designed to provide a large,
well-characterised sample of X-ray detected clusters suitable for
cosmological studies
\citep[][\citetalias{pierre2016}]{pierre2016}. The number counts of
the 100 brightest XXL clusters provided preliminary cosmological
hints: adopting the mass and temperature scaling relations
self-consistently measured from the same sample, \citet[][hereafter
  \citetalias{pacaud2016}]{pacaud2016} found a discrepancy with the
cluster density expected from the Planck 2015 cosmology
\citep{PlanckXIII2016}. This issue is quantitatively revisited with a
much larger sample in \citet[][\citetalias{pacaud2018}]{pacaud2018}.

We present here the first measurements of the 2PCF of XXL
clusters. With a statistical method designed to be independent of the
cluster selection function, we compare our measurements with standard
$\Lambda$-cold dark matter ($\Lambda$CDM) predictions, deriving
constraints on the total matter energy density parameter, $\Omega_{\rm
  M}$, and on the effective bias of the sample, $b_{eff}$.

All the numerical computations have been performed with the {\small
  CosmoBolognaLib}, a large set of {free software} libraries that
provide a highly optimised framework for managing catalogues of
extragalactic sources, measuring statistical quantities, and
performing Bayesian inferences on cosmological model parameters
\citep{marulli2016}\footnote{Specifically, we use the {\small
    CosmoBolognaLib} V4.1, which implements {\small OpenMP} parallel
  algorithms both to measure the 2PCF (Section \ref{sec:measurements})
  and to perform Markov chain Monte Carlo sampling (Section
  \ref{sec:cosmological}). The {\small CosmoBolognaLib} is entirely
  implemented in C++ to provide a high-performance back end for all
  the computationally expensive tasks, while also supporting their use
  in Python to exploit the higher-level abstraction of this
  language. The software and its documentation are freely available at
  the GitHub repository:
  \url{https://github.com/federicomarulli/CosmoBolognaLib} .}.

For consistency with the analyses presented in previous XXL papers, we
assume a fiducial $\Lambda$CDM cosmological model with WMAP9
parameters:  $\Omega_m = 0.28$, $\Omega_\Lambda = 0.72$,
$\Omega_b = 0.046$, $\sigma_8 = 0.817$, $n_s = 0.965$
\citep{hinshaw2013}. The dependence of observed coordinates on the
Hubble parameter is indicated as a function of $h\equiv H_0/100\, {\rm
  km\, s^{-1} Mpc^{-1}}$.

The paper is organised as follows. After the presentation of the XXL
cluster selection in Section \ref{sec:dataset}, we describe the
 methods adopted to measure and model the cluster clustering in Section
\ref{sec:measurements} and Section \ref{sec:cosmological},
respectively. We present and discuss our results in Section
\ref{sec:results}, and draw our main conclusions in Section
\ref{sec:conclusions}. Finally, Appendix \ref{sec:systematics}
provides a detailed investigation of the main systematics that might
impact the results presented in this work.


\section{Cluster sample}
\label{sec:dataset}

The catalogue analysed in this work is drawn from the second public
release of the XXL survey. The survey covers two extragalactic sky
regions of $\sim 50\,{\rm deg}^2$ in total, down to a point-source
sensitivity of $\sim 6\cdot10^{-15}
\mbox{erg}\,\mbox{s}^{-1}\,\mbox{cm}^{-2}$, in the [0.5--2] keV band
\citep[$90\%$ completeness
  limit;][\citetalias{chiappetti2018}]{chiappetti2018}. The data
processing pipeline and subsequent cluster detection (extended
sources) are described in detail in \citetalias{pacaud2016}; we
briefly summarise below the main steps.

In order to quantitatively deal with the completeness versus  purity issues
in the X-ray cluster selection process, we defined two samples of
extended sources in the [flux - apparent size] parameter space from
the X-ray pipeline output. This allowed us to compute accurate cluster
selection functions by means of extensive simulations.  The C1 class is
defined as having no contamination, that is no point sources
misclassified as extended. The C2 class corresponds to fainter, thus
less easily characterised, extended sources with an initial
contamination level of $\sim 50\%$, which is then a posteriori
eliminated by manual inspection of X-ray/optical overlays. We defined
a third class, C3, corresponding to (optical) clusters associated with
some X-ray emission, too weak to be characterised. Initially, most of
the C3 objects were not selected from the X-ray waveband and the
selection function of this subsample is undefined.

The current sample of spectroscopically confirmed extended X-ray
sources consists of $365$ galaxy clusters in total \citep[][hereafter
  \citetalias{adami2018}]{adami2018}. We considered the clusters
listed in \citetalias{adami2018}, for which we have a defined
measurement of $M_{500}$\footnote{$M_{500}$ is defined as the mass
  within radius $R_{500}$, which is the radius enclosing a mean
  density of $500$ times the critical density at the redshift of the
  cluster.}, which amounts to $182$ C1, $119$ C2, and $38$ C3
clusters. In this paper, we concentrate exclusively on the C1 sample;
however, in Appendix \ref{subsec:selection} we make a short digression
on the cosmological constraints from the 2PCF of C1+C2
clusters. Hence, all clusters analysed in the present study can be
considered as bona fide clusters: the C1 clusters constitute a
complete sample (in the cosmological sense), while the current C2
sample is pure but not yet complete \citepalias{adami2018}.

We usually estimate cluster masses by means of the mass-temperature
relation determined in \citet[][\citetalias{lieu2016}]{lieu2016}.
However, since not all C1 clusters have a temperature measurement, in
this article we rely on masses derived from a system of
self-consistent scaling relations. These relations are based on the
XMM count rates measured in an aperture of $300$ kpc (see Appendix F
of \citetalias{adami2018}).

The redshift and mass distributions of the XXL C1 and C2 clusters at
$z<1.5$ are shown in Figs. \ref{fig:nz} and \ref{fig:nM},
respectively. The C1 cluster catalogue considered in this work has a
median redshift $\langle z \rangle=0.317$ and a median mass $\langle
M_{500} \rangle\simeq1.3\cdot10^{14} M_\odot$.


\section{Measurements}
\label{sec:measurements}


\subsection{From observed to comoving coordinates}
\label{sebsec:map}

The first step required to measure the 2PCF is to convert observed redshifts
into distances. In standard cosmological frameworks the cosmological
redshifts, $z$, caused by the expansion of space, are related to
comoving distances, $d_{\rm c}$, as 
\begin{equation}
d_{\rm c} = c\int_0^z\frac{dz'}{H(z')} \; ,
\label{eq:distance}
\end{equation} 
where $c$ is the speed of light, and $H$ is the Hubble expansion
rate. Assuming a flat $\Lambda$CDM model, we have
\begin{equation}
H = H_0\left[\Omega_{\rm M}(1+z)^3+(1-\Omega_{\rm M})\right]^{1/2} \; .
\label{eq:Hubble}
\end{equation}
The observed redshift, $z_{\rm obs}$, is related to the cosmological
value by the  relation (neglecting redshift errors and
second-order corrections)
\begin{equation}
z_{\rm obs} = z + \frac{v_\parallel}{c}(1+z) \; ,
\label{eq:redshift}
\end{equation}
where $v_\parallel$ is the line-of-sight component of the
centre-of-mass velocity of the source. Since the peculiar velocities
are not directly measurable, we compute the comoving distances by
substituting $z$ with $z_{\rm obs}$ in Eq. \ref{eq:distance}. This
introduces distortions along the line of sight that are generally
called redshift-space distortions. Hereafter, we  refer to the
redshift-space spatial coordinates using the vector {\bf s}, whereas
we will use {\bf r} to indicate real-space coordinates. In the
following analysis we will neglect the measurement errors on the
observed redshifts since these are subdominant relative to the
clustering measurement uncertainties \citep[see e.g.][]{marulli2012b,
  sridhar2017}.

\begin{figure}
  \includegraphics[width=0.49\textwidth]{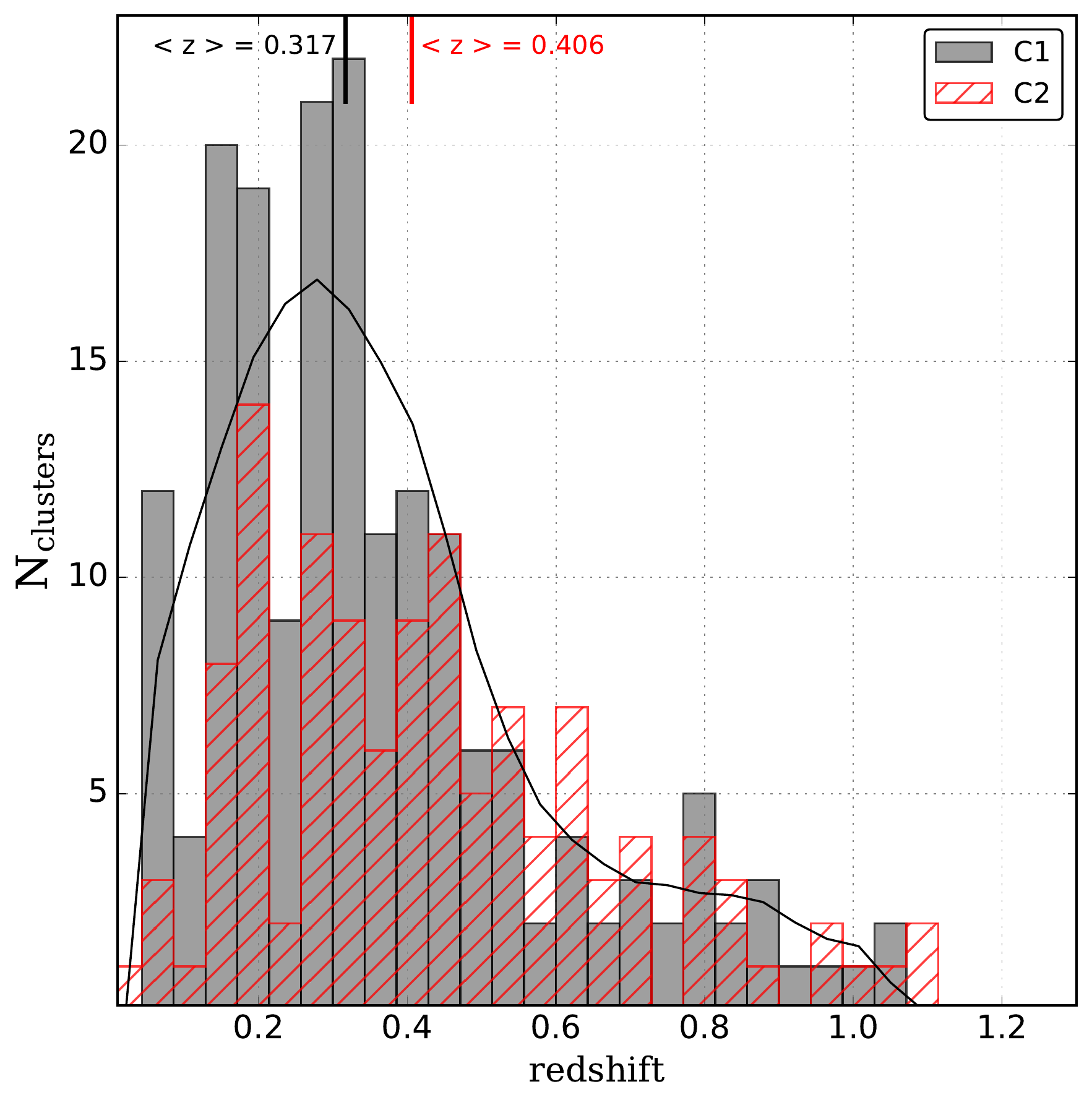}
  \caption{Redshift distribution of XXL C1 (grey histogram) and C2
    (red histogram) clusters at $z<1.5$, and of the C1 random objects
    normalised to the number of XXL C1 clusters (black line). The
    median redshifts of C1 and C2 clusters are shown at the top of the
    box.}
  \label{fig:nz}
\end{figure}


\subsection{ Two-point correlation function estimator}
\label{sebsec:twopoint}

An efficient way to investigate the large-scale structure of the
Universe is to compress its information content into the second-order
statistics of extragalactic sources, that is the 2PCF and power
spectrum \citep{totsuji1969, peebles1980}.

We measure the redshift-space angle-averaged 2PCF using the
\citet{landy1993} estimator,
\begin{equation}
  \hat{\xi}(s) \, = \, \frac{N_{RR}}{N_{CC}} \frac{CC(s)}{RR(s)} -2
  \frac{N_{RR}}{N_{CR}} \frac{CR(s)}{RR(s)} +1 \,,
  \label{eq:xiLS}
\end{equation}
where $CC(s)$, $RR(s),$ and $CR(s)$ are the binned numbers of
cluster-cluster, random-random, and cluster-random pairs with distance
$s \pm\Delta s$, while $N_{CC}=N_C(N_C-1)/2$, $N_{RR}=N_R(N_R-1)/2$,
and $N_{CR}=N_CN_R$ are the total numbers of cluster-cluster,
random-random, and cluster-random pairs in the sample, respectively
(see Appendix \ref{subsec:estimator} for more details).  It has been
demonstrated that the \citet{landy1993} estimator provides an unbiased
estimate of the 2PCF (in the limit $N_R\rightarrow\infty$), with
minimum variance. We define the comoving separation associated with
each bin as the average cluster pair separation inside the bin, which
is more accurate than using the bin centre, especially at large
scales, where the bin size is increasingly large
\citep[e.g.][]{zehavi2011}.

\begin{figure}
  \includegraphics[width=0.49\textwidth]{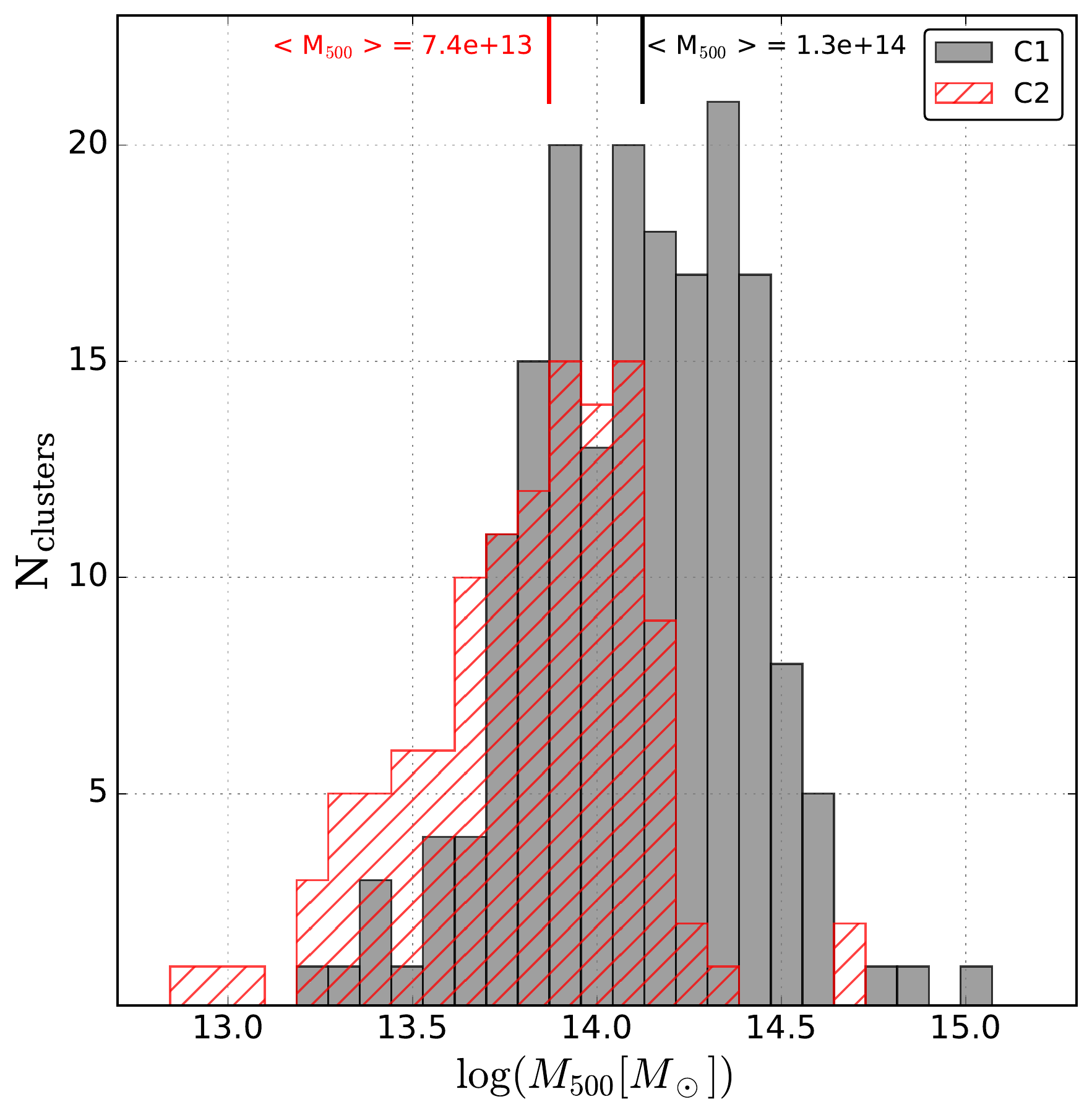}
  \caption{Mass distribution of XXL C1 (grey histogram) and C2 (red
    histogram) clusters at $z<1.5$. The median $M_{500}$ masses of C1
    and C2 clusters are shown at the top of the box.}
  \label{fig:nM}
\end{figure}


\subsection{Random catalogue}
\label{sebsec:random}

To estimate the 2PCF via Eq. \ref{eq:xiLS}, a random catalogue is
required, i.e. a catalogue of randomly distributed points having
the same three-dimensional coverage of the data. We adopt the common
assumption that the angular and redshift distributions of the tracers
are independent and can be treated separately. Following the same
methodology applied in \citet{gilli2005} and \citet[][hereafter
  \citetalias{plionis2018}]{plionis2018}, we construct the random
catalogue as follows.  We assign angular coordinates
(R.A.-Dec pairs) to the random objects by randomly extracting from
the real C1 XXL cluster coordinates, thus reproducing the same angular
distribution of the real sources. The redshifts are then assigned by
sampling from the Gaussian filtered radial distribution of the XXL
clusters \citep[e.g.][]{marulli2013}. The smoothing is necessary to
avoid spurious clustering along the line of sight. We set the
smoothing length to $\sigma_z=0.1$ \citepalias[see][for further
  details]{plionis2018}. The redshift distribution of the random
objects normalised to the number of XXL C1 clusters is shown in
Fig. \ref{fig:nz}.

This method has the advantage of relying  on real angular
coordinates alone, without any need to model the angular mask.  Different
assumptions on the angular selection used to construct the random
catalogue might impact the measurement at small scales
($s\lesssim10$\Mpch), though the effect is within the current
measurement uncertainties (see Appendix \ref{subsec:random}). To
minimise the impact of shot noise error due to the finite number of
random objects, we construct our random catalogue to be $100$ times
larger than the XXL cluster sample.


\subsection{Covariance matrix}
\label{sebsec:covariance}

The covariance matrix, $C_{i,j}$, which measures the variance and
correlation between 2PCF bins, is defined as 
\begin{equation}
  C_{i,j} = \mathcal{F} \sum_{k=1}^{N_R}(\xi^k_i-\bar{\xi_i})
  (\xi^k_j-\bar{\xi_j}) \, ,
  \label{eq:covariance}
\end{equation}
where the subscripts $i$ and $j$ run over the 2PCF bins, $k$ refers to
the 2PCF of the $k^{th}$ of $N_R$ catalogue realisations, and
$\bar{\xi}$ is the mean 2PCF of the $N_R$ samples. The normalisation
factor, $\mathcal{F}$, which takes into account the fact that the
$N_R$ realisations might not be independent, is $\mathcal{F}=1/N_R$,
$\mathcal{F}=(N_R-1)/N_R$ and $\mathcal{F}=1/(N_R-1)$ for the subsample,
jackknife, and bootstrap methods, respectively \citep{norberg2009}.  We
assess the XXL 2PCF covariance matrix with the bootstrap method, using
$1000$ realisations obtained by resampling galaxy clusters from the
original catalogue, with replacement. The impact of this choice is
discussed in Appendix \ref{subsec:covariance}.


\section{Cosmological analysis}
\label{sec:cosmological}

We perform a Bayesian statistical Markov chain Monte Carlo (MCMC)
analysis of the 2PCF by sampling the posterior distribution of
$\Omega_{\rm M}$, the only free parameter in the assumed flat
$\Lambda$CDM model considered. As we verified, the current clustering
uncertainties do not allow us to consider more general cosmological
scenarios, for example models with free dark energy equation of state
parameters, by exploiting only the XXL cluster clustering. The joint
cosmological analysis of XXL cluster number counts and clustering will
be presented in a forthcoming paper.

We consider the commonly used likelihood function, $\mathscr{L}$,
defined as 
\begin{equation}
  -2\ln\mathscr{L} = \sum_{i=1}^{N}\sum_{j=1}^{N}
  (\xi_i^{d}-\xi^{m}_i)C^{-1}_{i,j}(\xi_j^{d}-\xi^{m}_j) \, ,
\label{eq:likelihood}
\end{equation}
where $C^{-1}_{i,j}$ is the inverse of the covariance matrix estimated
from the data with the bootstrap method (Eq. \ref{eq:covariance}), $N$
is the number of comoving separation bins at which the 2PCF ($\xi$) is
estimated, and the superscripts $d$ and $m$ stand for data and
model. The likelihood is estimated at the mean pair separations in
each bin (see Section \ref{sebsec:twopoint}).

The 2PCF model in redshift space, $\xi^{m}(s)$, is computed as

\begin{equation}
  \xi^{m}(s) = \left[ (b_{eff}\sigma_8)^2 + \frac{2}{3} f\sigma_8
    \cdot b_{eff}\sigma_8 + \frac{1}{5}(f\sigma_8)^2 \right]
  \frac{\xi_{\rm DM}(\alpha r)}{\sigma_8^2} \; ,
  \label{eq:xi0} 
\end{equation}
where $\xi_{\rm DM}(r, z)$ is the real-space dark matter 2PCF, which we
estimated by Fourier transforming the power spectrum, $P_{\rm DM}(k,
z)$, computed with the software {\small CAMB} \citep{lewis2002}. 
%
Since the present cluster clustering analysis focuses at sufficiently
large scales, the dark matter power spectrum can be safely computed in
linear theory,   $P_{\rm DM}(k, z)\simeq P_{\rm DM}^{lin}(k,
z)$, with marginal effects on our results. The model depends on two
free quantities, $f\sigma_8$ and $b_{eff}\sigma_8$ (since $\xi_{\rm
  DM}\propto\sigma_8^2$), and on the reference background cosmology
used both to convert angles and redshifts into distances and to
estimate the real-space dark matter 2PCF \citep[see
  e.g.][]{marulli2017}. The geometric distortions caused by an
incorrect assumption of the background cosmology are modelled by the
$\alpha$ parameter, i.e. the ratio between the test and fiducial
values of the isotropic volume distance, $D_V$, defined as 
\begin{equation}
  D_V \equiv \left[(1+z)^2D_A^2\frac{cz}{H}\right]^{1/2} \; ,
  \label{eq:Dv}
\end{equation}
where $D_A$ is the angular diameter distance
\citep[][]{eisenstein2005}. The $\alpha$ parameter allows us to fit
the 2PCF estimated with the fiducial cosmological model without the
need to re-measure it for every cosmological model tested in the MCMC.

Equation \ref{eq:xi0} provides a mapping from real space to redshift
space in the distant-observer approximation, assuming that non-linear
redshift-space distortion effects can be neglected \citep{kaiser1987,
  lilje1989, mcgill1990, hamilton1992, fisher1994}. This is a
reasonable assumption in order to model the clustering of galaxy
clusters at the scales considered in this analysis; in other words,
the impact of neglecting the Fingers of God effect is marginal,
considering current measurement uncertainties \citep[see][and
  references therein]{marulli2017}. The $f$ and $b_{eff}$ parameters
in Eq. \ref{eq:xi0} are the linear growth rate and the linear
effective bias of the sample, respectively. Specifically, $f\equiv
d\log\delta/d\log a$, where $a$ is the dimensionless scale factor and
$\delta$ is the growing mode linear fractional density
perturbation. It can be approximated as $f(z) \simeq \Omega_{\rm
  M}^{\gamma}(z)$ in most cosmological scenarios, with
$\gamma\simeq0.545$ in $\Lambda$CDM \citep{wang1998, linder2005}.

One of the great advantages of using galaxy clusters (instead of galaxies) as density
tracers  is that we can have an estimate of their
total masses, and thus predict their bias given an assumed
cosmological model. Following a similar approach to
\citet{moscardini2000}, we account for light-cone effects by
estimating the effective bias as the average over the selected cluster
pairs, 
\begin{equation}
  b_{eff}^2 = <b(\tilde{M}_i,z_i)b(\tilde{M}_j,z_j)> \; ,
\label{eq:bias_eff}
\end{equation}
where $\tilde{M_i}$ and $\tilde{M_j}$ are the masses of the two XXL
clusters of each pair at redshift $z_i$ and $z_j$, respectively,
assessed by sampling from a Gaussian distribution with standard
deviation equal to the given mass uncertainty (see Section
\ref{sec:dataset} for details). The linear bias of each cluster, $b$,
is computed with the \citet{tinker2010} model for $M_{500}$.

\begin{figure}
  \includegraphics[width=0.49\textwidth]{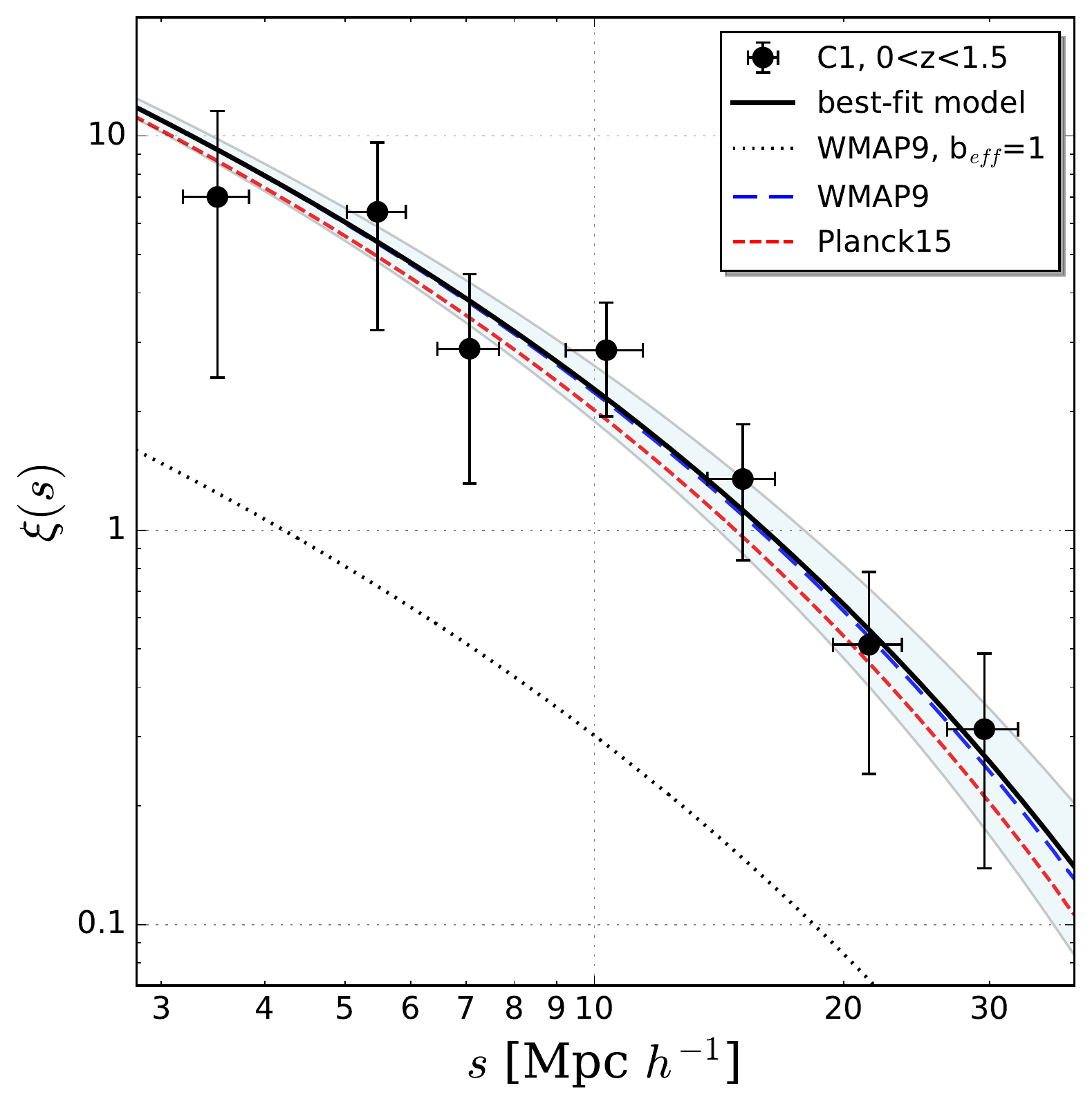}
  \caption{Redshift-space 2PCF of the C1 XXL clusters at $z<1.5$
    (black dots) compared to the best-fit model, i.e. the median of
    the MCMC posterior distribution (black solid line). The shaded
    area shows the $68\%$ uncertainty on the posterior median. The
    derived best-fit model correlation length is $s_0=16\pm2$
    \Mpch. The red dashed and blue long-dashed lines show the WMAP9
    and Planck15 predictions, respectively, computed as described in
    Sect. \ref{sec:cosmological}. Their correlation lengths are
    $s_0=15.83$ \Mpch and $s_0=14.81$ \Mpch, respectively.  The
    vertical error bars are the diagonal values of the bootstrap
    covariance matrix, while the horizontal error bars show the
    standard deviation around the mean pair separation in each
    bin. The black dotted line shows the WMAP9 prediction with
    $b_{eff}=1$ as a reference.}
  \label{fig:xi}
\end{figure}


\section {Results}
\label{sec:results}


\subsection{ XXL cluster clustering}
\label{subsec:clustering}

\begin{figure}
  \includegraphics[width=0.49\textwidth]{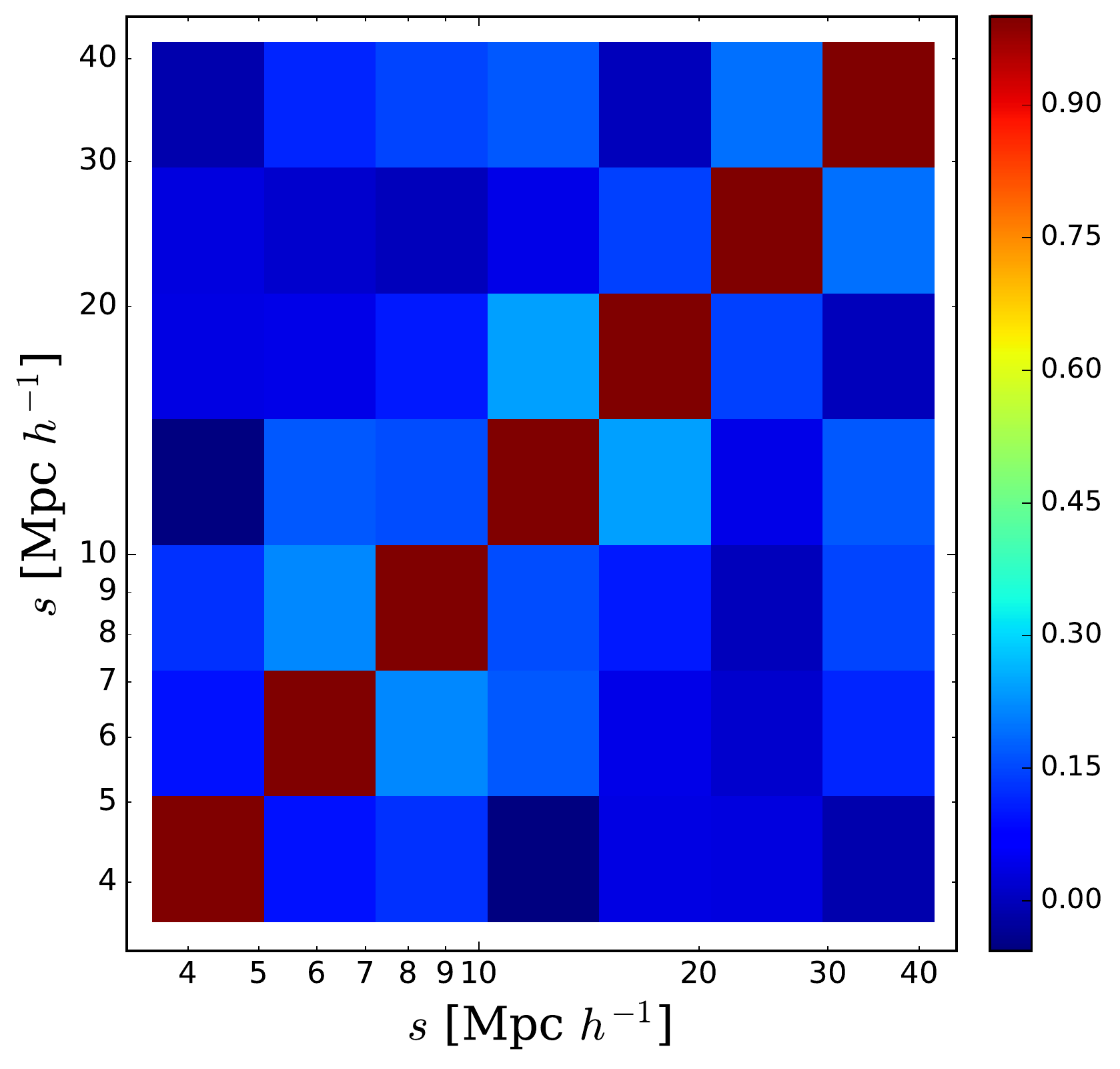}
  \caption{Bootstrap correlation matrix
    $\left(C_{i,j}/\sqrt{C_{i,j}C_{j,i}}\right)$ of C1 XXL clusters at
    $z<1.5$.}
  \label{fig:Bootstrap}
\end{figure}

Figure \ref{fig:xi} shows the redshift-space 2PCF of the C1 XXL
clusters at $z<1.5$. The clustering function is measured, as described
in Section \ref{sebsec:twopoint}, in eight equal logarithmic bins in
the comoving separation range $3<s\,[$\Mpch$]<50$. At smaller
separations the clustering signal is not detectable in our data, due
to the minimum cluster separation set by the cluster sizes and to the
density of the catalogue. On the other hand, at scales larger than
those shown in Fig. \ref{fig:xi} the signal is dominated by the sample
variance, due to the XXL volume. The vertical error bars are the
diagonal values of the bootstrap covariance matrix (see Section
\ref{sebsec:covariance}), while the horizontal error bars represent
the standard deviation around the mean pair separation in each
bin. The full bootstrap correlation matrix, defined as
$C_{i,j}/\sqrt{C_{i,j}C_{j,i}}$ (see Eq. \ref{eq:covariance}), is
shown in Fig. \ref{fig:Bootstrap}.


\subsection{Constraints on $\Omega_{\rm M}$ and $b_{eff}$}
\label{subsec:constraints}

We model the XXL cluster clustering following the statistical method
described in Section \ref{sec:cosmological}. Figure \ref{fig:xi}
compares the XXL 2PCF to the best-fit model. Specifically, we show the
posterior MCMC median, together with the $68\%$ uncertainty around the
median.  The fitting analysis is performed in the scale range
$10<r[$\Mpch$]<40$, where the signal is robust (see Section
\ref{subsec:clustering}), though the final results are marginally
affected by this choice (see Appendix \ref{subsec:fitting}). The
likelihood function is constructed by assuming a flat $\Lambda$CDM
model with one free parameter $\Omega_{\rm M}$, for which we assume a
flat prior in the range $[0, 1]$. All the other parameters are set to
WMAP9 values, with a Gaussian prior on $\sigma_8$ with mean $0.817$
and standard deviation $0.02$. The effective bias is a derived
parameter that is updated at each MCMC step.

\begin{figure}
  \includegraphics[width=0.51\textwidth]{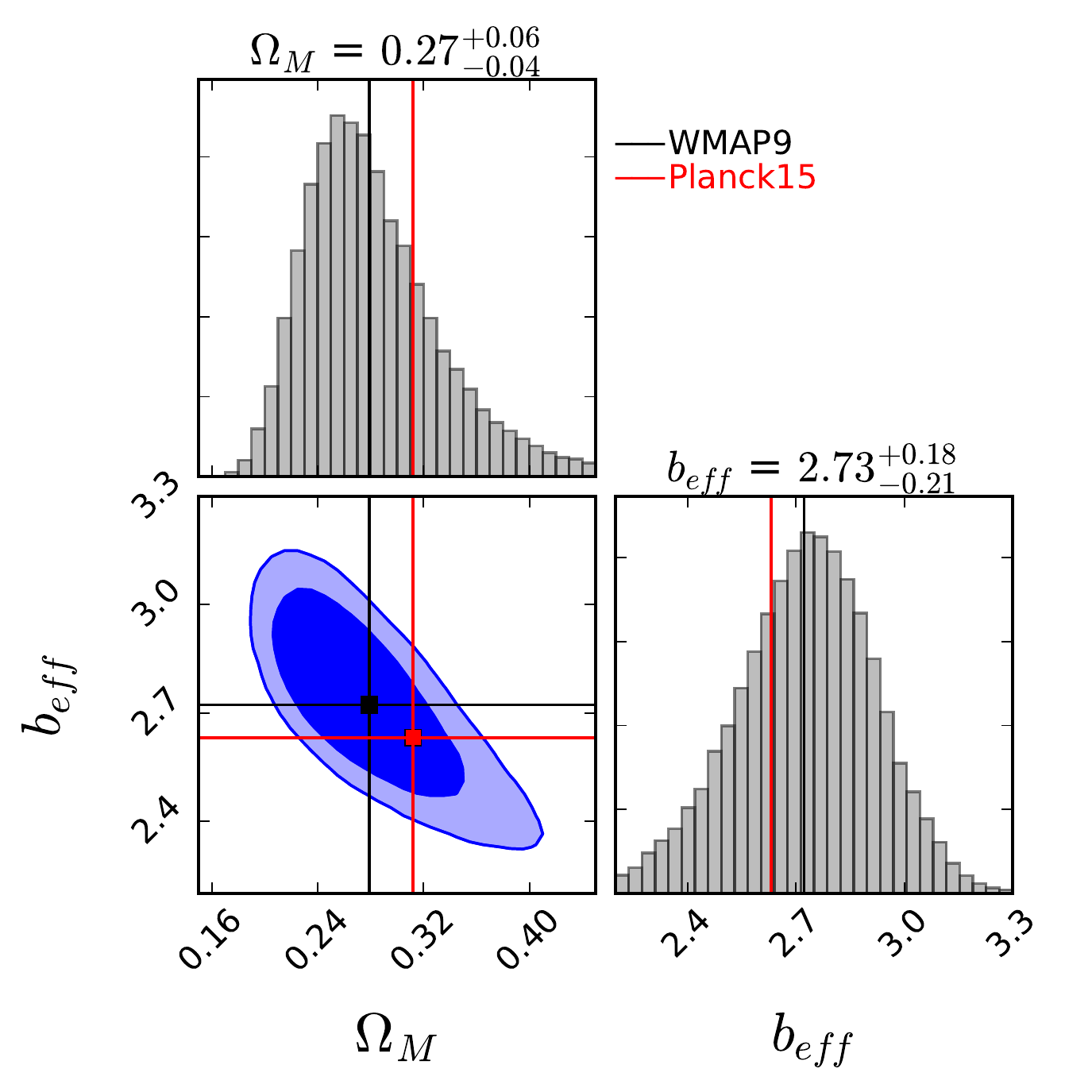}
  \caption{Confidence contours ($1-2\sigma$) of $\Omega_{\rm
      M}-b_{eff}$ provided by the MCMC, as described in Sect.
    \ref{sec:cosmological} ($b_{eff}$ is a derived parameter, the
    ellipse width corresponds to the deviation of the $\sigma_8$
    Gaussian prior). The histograms (top and bottom right panels) show
    the posterior distributions of $\Omega_{\rm M}$ and $b_{eff}$,
    respectively. Black and red lines represent WMAP9 and Planck15
    predictions, respectively.}
  \label{fig:constraints}
\end{figure}

The $1-2\sigma$ confidence contours of $\Omega_{\rm M}-b_{eff}$
provided by the MCMC are shown in Fig. \ref{fig:constraints}. We obtain
$\Omega_{\rm M}=0.27_{-0.04}^{+0.06}$ and
$b_{eff}=2.73_{-0.20}^{+0.18}$, where the best-fit values are the MCMC
medians, while the errors are estimated as the $1-\sigma$ of the
posterior probability distribution. As expected, this distribution is
not symmetric about the mode. The derived best-fit model correlation
length, that is the scale for which the redshift-space 2PCF is equal
to $1$, is $s_0=16\pm2$ \Mpch.

Our measurements appear fully consistent with $\Lambda$CDM
predictions, in agreement with previous cosmological XXL analyses
\citepalias{pacaud2016, adami2018, pacaud2018}, providing a new and
independent confirmation of the standard cosmological framework.
Moreover, Fig. \ref{fig:xi} demonstrates that the effective bias
estimated from the XXL cluster masses via Eq. \ref{eq:bias_eff} is
consistent with what is expected to match the measured clustering
normalisation.

The XXL clustering uncertainties are still too large, however, to
allow us to discriminate between WMAP9 and Planck15 cosmologies: both
appear consistent with the data.  This is shown in Fig. \ref{fig:xi},
where the measured XXL cluster clustering is compared to the
theoretical correlation function computed with
Eqs.\ref{eq:xi0}--\ref{eq:bias_eff}, assuming WMAP9 \citep{hinshaw2013}
and Planck15 \citep{PlanckXIII2016} cosmological parameters, as
provided in their Table 4 (TT, TE, EE+lowP+lensing), that is $\Omega_m
= 0.3121$, $\Omega_\Lambda = 0.6879$, $\Omega_b = 0.0488$, $\sigma_8 =
0.8150$, $n_s = 0.9653$. The effective bias values predicted by the
\citet{tinker2010} model in WMAP9 and Planck15 cosmologies, respectively
$b_{eff}=2.72$ and $b_{eff}=2.63$,  are indicated by lines
in Fig. \ref{fig:constraints}. The correlation lengths of WMAP9 and
Planck15 cosmologies are $s_0=15.83$ \Mpch and $s_0=14.81$ \Mpch,
respectively.

Clustering and number counts provide independent complementary probes
that can be combined together. As Figs. \ref{fig:xi} and
\ref{fig:constraints} demonstrate, this is indeed feasible with XXL
data, as the 2PCF signal-to-noise ratio is sufficient at the scales
shown. This issue will be addressed in a forthcoming work.


\section {Conclusions}
\label{sec:conclusions}

We investigated the spatial properties of the largest homogeneous
survey of X-ray selected galaxy clusters to date, carried out by the
{\em XMM-Newton} satellite, and compared the measurements with
standard $\Lambda$CDM predictions. The main results of this analysis
are summarised below:
\begin{itemize}
\item
  We measured the 2PCF in redshift space of a sample of $182$ X-ray
  selected galaxy clusters at median redshift $\langle z
  \rangle=0.317$ and median mass $\langle M_{500}
  \rangle\simeq1.3\cdot10^{14} M_\odot$. This is the first time that
  the clustering of an X-ray selected cluster catalogue at such
  relatively high redshifts and low masses has been measured.
\item
  We modelled the data by performing an MCMC analysis, assuming a flat
  $\Lambda$CDM cosmology. Exploiting the XXL cluster clustering
  measurements in combination with cluster mass estimates from scaling
  relations, used to derive the effective bias, we implemented a
  statistical method independent of the cluster selection function.
\item
  We found that the 2PCF of XXL clusters is consistent with the
  $\Lambda$CDM predictions. We obtain $\Omega_{\rm
    M}=0.27_{-0.04}^{+0.06}$ and $b_{eff}=2.73_{-0.20}^{+0.18}$. The
  derived redshift-space correlation length of the C1 XXL clusters is
  $s_0=16\pm2$ \Mpch. This provides an important confirmation of the
  standard model, which is independent of the cluster number counts
  and of the other standard cosmological probes, such as  the
  galaxy clustering.
\item
  This work also demonstrates that the effective linear bias computed
  from cluster masses estimated with scaling relations is consistent
  with the expected cluster clustering normalisation.
\item
  Though the current measurement uncertainties are not small enough to
  discriminate between WMAP9 and Planck15 cosmologies, this work
  demonstrates the feasibility of a cosmological exploitation of XXL
  cluster clustering, paving the way for a joint analysis in
  combination with cluster number counts.
\end{itemize}
The combination of cluster number counts and clustering is especially
powerful when the dark energy equation state parameter is left
free. This will thus allow us to constrain a much wider parameter
space, as already attempted in \citetalias{pacaud2018} with number
counts alone. Moreover, in the final combined analysis, we will use
the full C1+C2 cluster sample.

The next generation of galaxy surveys, such as the Dark Energy
Survey\footnote{\url{http://www.darkenergysurvey.org}} (DES)
\citep{DES2017}, the extended Roentgen Survey with an Imaging
Telescope Array (eROSITA) satellite
mission\footnote{\url{http://www.mpe.mpg.de/eROSITA}}
\citep{merloni2012}, the NASA Wide Field Infrared Space Telescope
(WFIRST) mission\footnote{\url{http://wfirst.gsfc.nasa.gov}}
\citep{spergel2013}, the ESA Euclid
mission\footnote{\url{http://www.euclid-ec.org}} \citep{laureijs2011,
  amendola2016}, and the Large Synoptic Survey
Telescope\footnote{\url{http://www.lsst.org}} (LSST)
\citep{ivezic2008} will provide increasingly large catalogues of
galaxy clusters, extending the current redshift and mass ranges, and
eventually providing substantially tighter constraints on cosmological
parameters \citep{borgani2001, angulo2005, sartoris2016}.


\begin{acknowledgements}
XXL is an international project based around an XMM Very Large
Programme surveying two $25\,\mbox{deg}^2$ extragalactic fields at a
depth of
$\sim6\times10^{-15}\,\mbox{erg}\,\mbox{cm}^{-2}\,\mbox{s}^{-1}$ in
the [0.5--2] keV band for point-like sources. The XXL website is
\url{http://irfu.cea.fr/xxl}. Multi-band information and spectroscopic
follow-ups of the X-ray sources are obtained through a number of
survey programmes, summarised at
\url{http://xxlmultiwave.pbworks.com/}.

We thank L. Chiappetti and M. Roncarelli for helpful
comments. F.M. acknowledges support from the grant MIUR PRIN 2015
``Cosmology and Fundamental Physics: illuminating the Dark Universe
with Euclid''.  The Saclay group acknowledges long-term support from
the Centre national d'{\'e}tudes spatiales (CNES). S.E. acknowledges
financial contribution from the contracts NARO15 ASI-INAF I/037/12/0,
ASI 2015-046-R.0, and ASI-INAF n.2017-14-H.0. S.A. acknowledges support
from Istanbul University with the project number BEK-54547.
\end{acknowledgements}


\appendix

\section{Systematics}
\label{sec:systematics}

This section presents a detailed investigation of all the main
systematics that might impact the results of this work. In
\S\ref{subsec:selection} we test the impact of the sample
selection. In \S\ref{subsec:estimator} and \S\ref{subsec:covariance}
we investigate the estimators used in this work to measure the 2PCF
and assess its covariance matrix, respectively. In
\S\ref{subsec:random} we test the method used to construct the random
catalogue. In \S\ref{subsec:fitting} we discuss the impact of our
modelling assumptions. Finally, in \S\ref{subsec:masses} we
investigate the effect of mass uncertainties.


\subsection{Sample selection}
\label{subsec:selection}
The analysis presented in this work was performed using the full
sample of $182$ C1 XXL clusters at $z<1.5$. Here we investigate the
impact of this assumption.

\begin{figure}
  \includegraphics[width=0.49\textwidth]{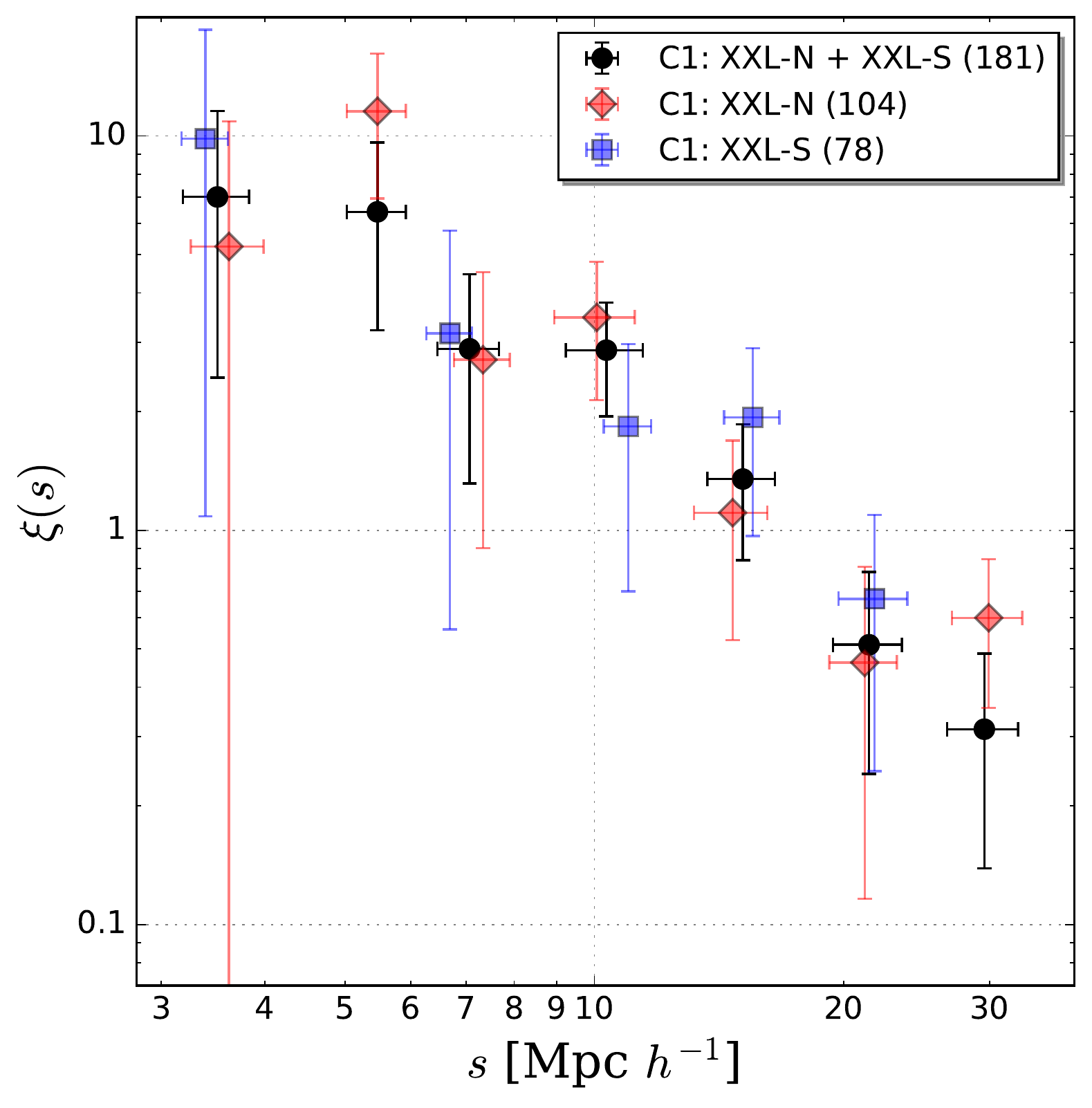}
  \caption{Comparison between the redshift-space 2PCF of XXL C1 in
    XXL-N (red diamonds), in XXL-S (blue squares), and in the whole
    sample (black dots). The number of XXL clusters in each field is
    reported in parentheses. The error bars are as in
    Fig. \ref{fig:xi}.}
  \label{fig:test1}
\end{figure}

Figure \ref{fig:test1} compares the redshift-space 2PCF of C1 XXL
clusters in the XXL-N and XXL-S fields separately, and in the whole
sample. Given the estimated errors, the three measurements appear
consistent with each other; there are no systematic differences.

\begin{figure}
  \includegraphics[width=0.49\textwidth]{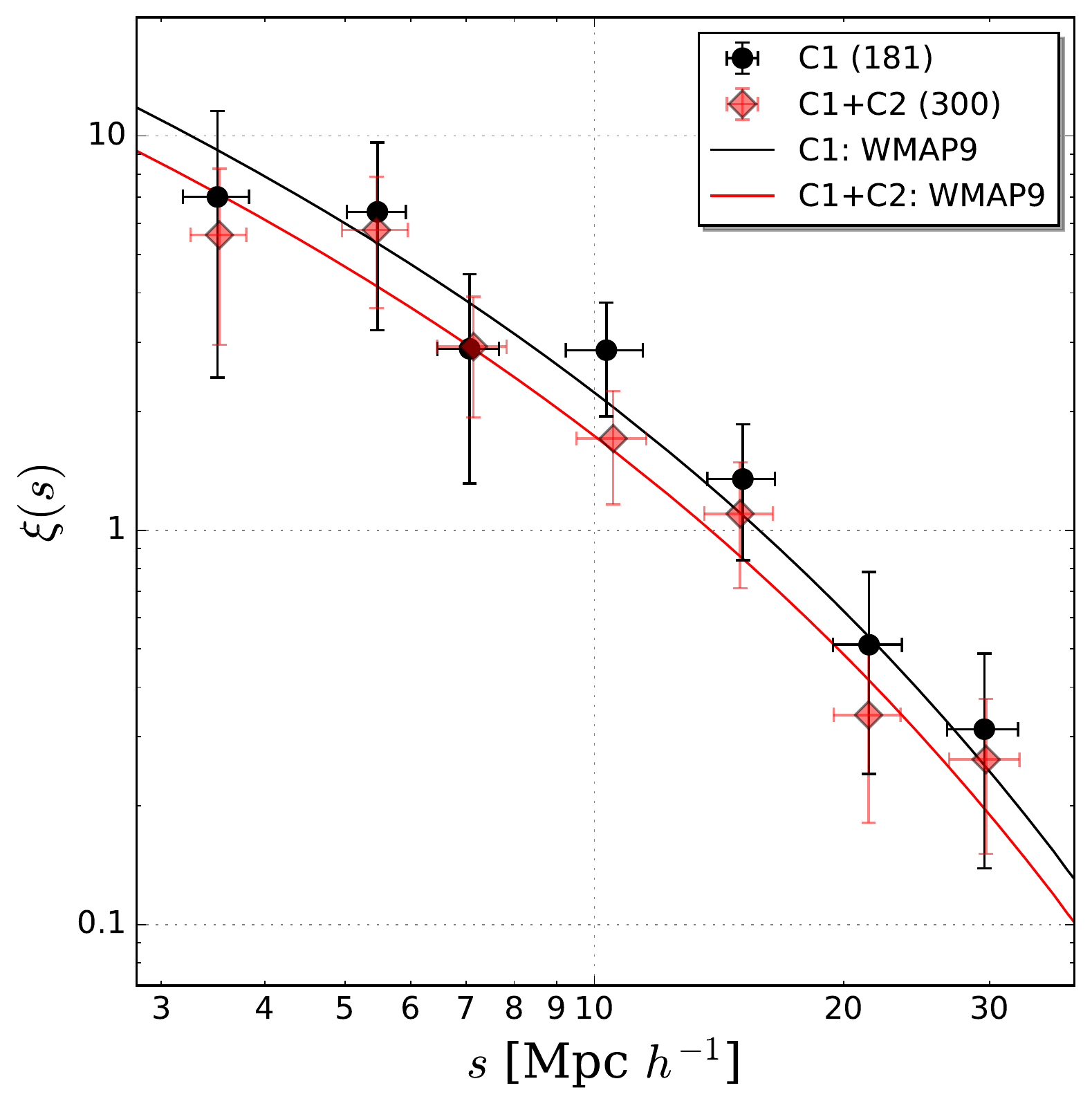}
  \caption{Comparison between the redshift-space 2PCF of XXL C1 (black
    dots) and C1+C2 clusters (red diamonds). The black and red lines
    show the theoretical WMAP9 predictions, computed as described in
    Sect. \ref{sec:cosmological} for C1 and C1+C2, respectively. The
    number of C1 and C1+C2 XXL clusters in each field is reported in
    parentheses.}
  \label{fig:test2}
\end{figure}

\begin{figure}
  \includegraphics[width=0.49\textwidth]{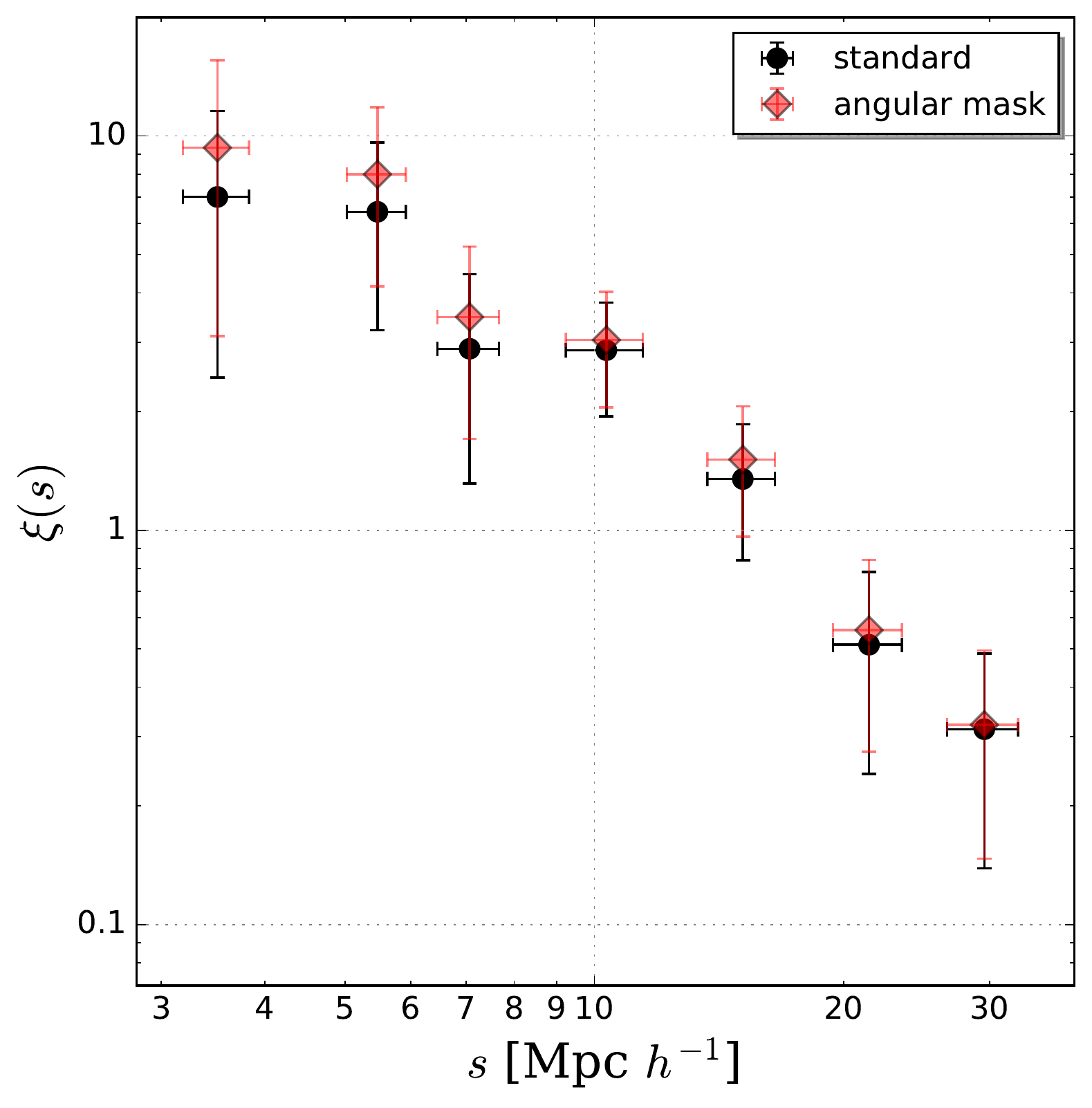}
  \caption{Comparison between the redshift-space 2PCF of XXL C1
    clusters computed with the random catalogue constructed as
    described in Sect. \ref{sebsec:random} (black dots) and
    considering the angular mask (red diamonds).}
  \label{fig:test4}
\end{figure}

Figure \ref{fig:test2} shows the 2PCF of the sample comprising both C1
and C2 XXL clusters. As expected, the clustering bias is lower than
that for C1 clusters as the mass distribution of the C2 sample is
shifted to lower masses (see Fig. \ref{fig:nM}). Due to the low
comoving number density of the C2 cluster sample, the C2 2PCF
measurement is highly uncertain, thus limiting our analysis to the
comparison between C1 and C1+C2 2PCFs. As shown in
Fig. \ref{fig:test2}, the 2PCFs of both the C1 and C1+C2 cluster
samples are found to be fully consistent with WMAP9 predictions. The
$1\sigma$ MCMC confidence contours of $\Omega_{\rm M}-b_{eff}$ are
shown in Fig. \ref{fig:test4}. We obtain $\Omega_{\rm
  M}=0.29_{-0.04}^{+0.05}$ and $b_{eff}=2.37_{-0.15}^{+0.14}$. As
expected, the errors on \Om and $b_{eff}$ are slightly smaller than
those obtained from the clustering of C1 clusters. To be conservative,
we decided to focus the analysis on the C1 sample, which is complete,
as discussed in Section \ref{sec:dataset}.

\begin{figure}
  \includegraphics[width=0.49\textwidth]{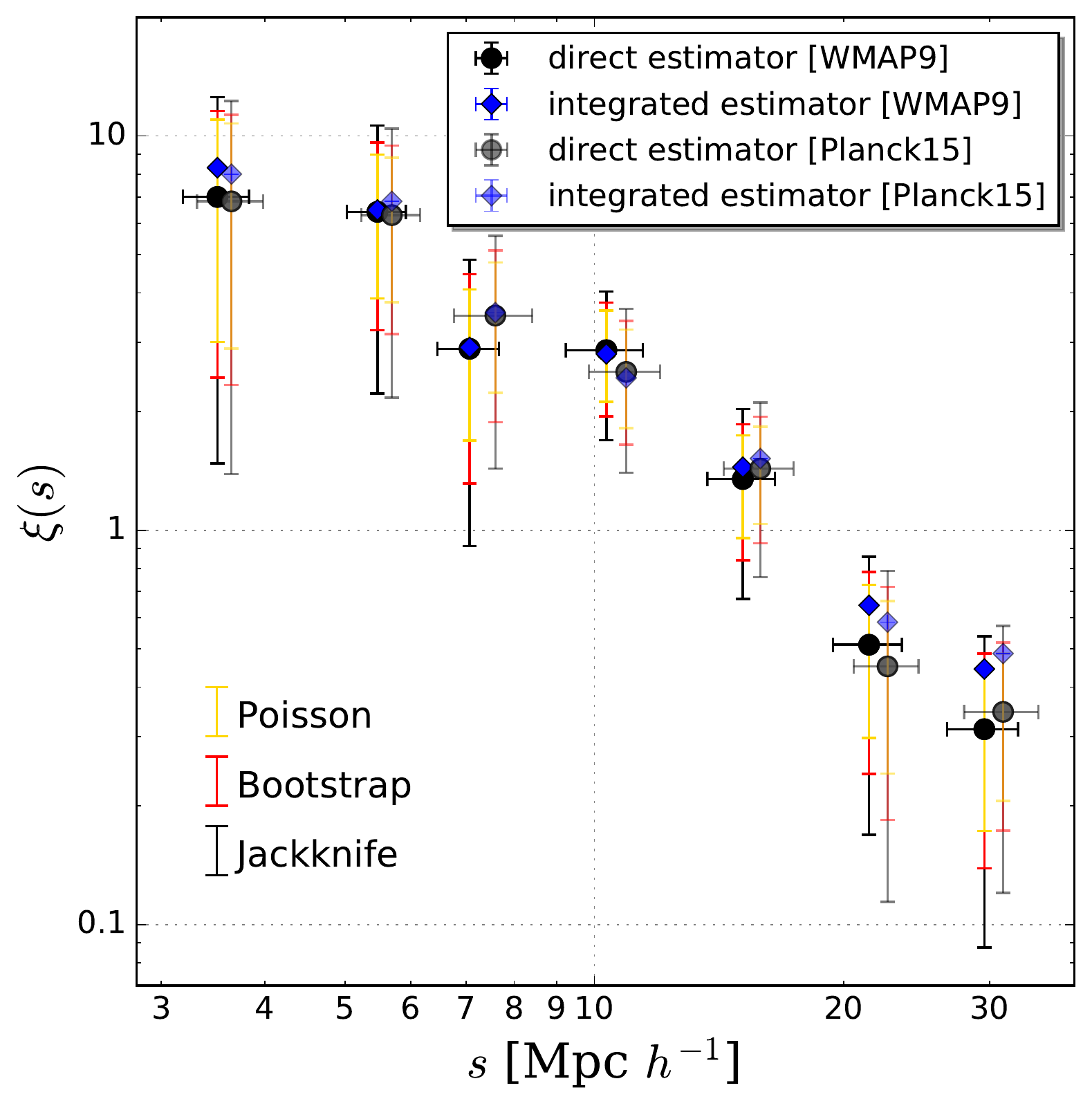}
  \caption{Comparison between the redshift-space 2PCF of XXL C1
    clusters computed with the direct (dots) and integrated (diamonds)
    estimators, assuming either WMAP9 (solid coloured) or Planck15
    (fuzzy coloured, slightly shifted for reasons of clarity). The
    error bars compare the Poisson, bootstrap, and jackknife estimated
    errors.}
  \label{fig:test3}
\end{figure}

\subsection{Clustering estimator}
\label{subsec:estimator}

The $\ell_{th}$ multipole of the 2PCF is defined as 
\begin{equation}
  \xi_\ell(r) = \frac{2\ell+1}{2}\int_{-1}^1\mbox{d}\mu\,
  L_\ell(\mu)\xi(r,\mu) \, ,
  \label{eq:integrated}
\end{equation}
where $L_\ell(\mu)$ is the $\ell_{th}$ Legendre polynomial and $\mu$
is the cosine of the angle between the galaxy separation and the line
of sight. The clustering monopole measured in this work corresponds to
$\ell=0$. The signal-to-noise ratio in the higher-order multipoles of the
XXL cluster clustering is not sufficient to allow any significant
statistical analysis. The clustering multipoles can be computed with
the \citet{landy1993} estimator as follows:
\begin{multline}
  \hat{\xi}_\ell(r) = \\ \frac{2\ell+1}{2}\int_{-1}^1\mbox{d}\mu\,
  L_\ell(\mu) \left( \frac{N_{RR}}{N_{CC}} \frac{CC(s, \mu)}{RR(s,
    \mu)} -2 \frac{N_{RR}}{N_{CR}} \frac{CR(s, \mu)}{RR(s, \mu)}
  +1\right) \, .
  \label{eq:integratedLS}
\end{multline}
This is the {\em integrated estimator} of the 2PCF multipoles. As
discussed in Section \ref{sebsec:twopoint}, the clustering
measurements presented in this work have been estimated with a
different {\em direct estimator} given by Eq. \ref{eq:xiLS}, which
computes pair counts in 1D scale bins directly.  The two estimators
coincide when the random pairs do not depend on $\mu$, that is
$RR(r,\mu)=RR(r)$ \citep{kazin2012}. This condition is verified in our
case, as demonstrated by Fig. \ref{fig:test3}, which shows that the
2PCFs measured with the integrated and direct estimators are
consistent.

The geometric distortions are modelled by the $\alpha$ parameter (see
Eq. \ref{eq:xi0}), though they are negligible given the estimated
uncertainties \citep{marulli2012b}. The 2PCFs measured assuming either
WMAP9 or Planck15 cosmologies are compared in Fig. \ref{fig:test3}.  We
find no significant differences between these measurements to within
the estimated errors of the 2PCF. Specifically, we obtain
$\alpha=1.003_{-0.017}^{+0.013}$ and $\alpha=1.012_{-0.019}^{+0.014}$,
assuming WMAP9 and Planck15 cosmologies, respectively.


\subsection{Covariance matrix}
\label{subsec:covariance}

As described in Section \ref{sebsec:covariance}, to estimate the XXL
2PCF covariance matrix we used the bootstrap method with $1000$
realisations. This number is large enough to assure convergence, as we
verified. We compare here this covariance matrix with that obtained
with the jackknife method, consisting in subsampling the original
catalogue and calculating the 2PCF in all but one subsample. In
particular, we apply this procedure by removing each cluster
recursively. The error bars shown in Fig. \ref{fig:test3} compare the
diagonal values of the jackknife and bootstrap covariance matrices. We
also show the estimated Poissonian errors for comparison. The
$1\sigma$ MCMC confidence contours of $\Omega_{\rm M}-b_{eff}$
obtained with the jackknife method are shown in
Fig. \ref{fig:test4}. In this case, we obtain $\Omega_{\rm
  M}=0.29_{-0.05}^{+0.09}$ and $b_{eff}=2.68_{-0.27}^{+0.21}$, fully
in agreement with the results obtained with bootstrap. We adopted the
latter as the reference as the XXL bootstrap covariance matrix is
smoother thanks to the larger number of possible resamplings.


\subsection{Random catalogue}
\label{subsec:random}

\begin{figure}
  \includegraphics[width=0.49\textwidth]{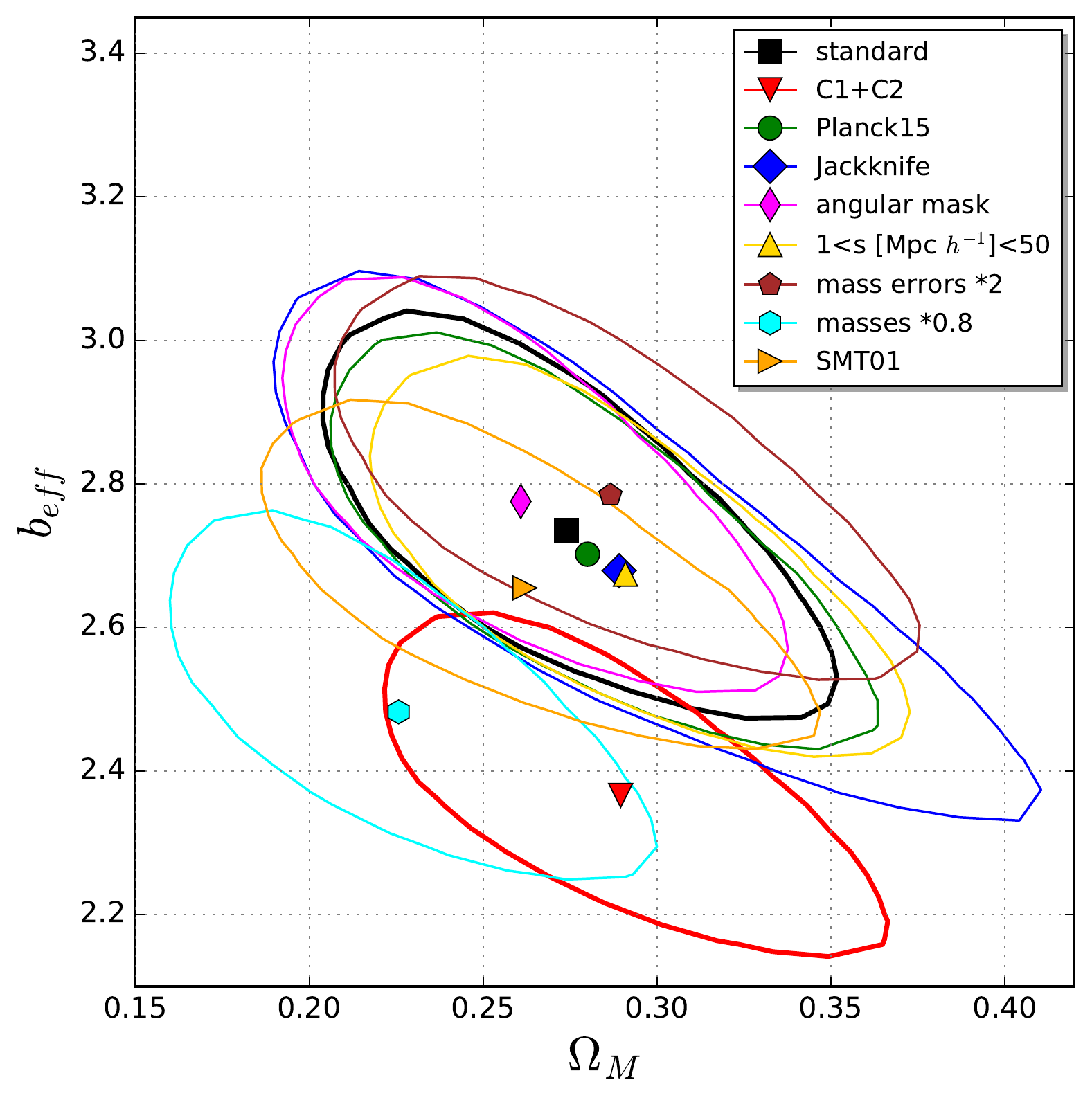}
  \caption{$1\sigma$ MCMC confidence contours of $\Omega_{\rm
      M}-b_{eff}$ obtained with different assumptions: standard
    analysis, as described in Sect. \ref{sec:cosmological} - black;
    considering C1+C2 XXL clusters, instead of C1 only - red; assuming
    Planck15 as reference cosmology, instead of WMAP9 - green; with
    jackknife covariance, instead of bootstrap - blue; considering the
    angular mask to construct the random catalogue, instead of the
    technique described in Sect. \ref{sebsec:random} - magenta;
    considering the fitting scale range $1<r[$\Mpch$]<50$, instead of
    $10<r\,[$\Mpch$]<40$ - yellow; doubling the statistical mass
    errors - brown; reducing the masses by $20\%$ - cyan; assuming the
    \citet{sheth_mo_tormen2001} bias model to compute the effective
    bias of the sample - orange.}
  \label{fig:test5}
\end{figure}

We test here the impact of the technique adopted to construct the
random catalogue. Figure \ref{fig:test4} compares the reference 2PCF
with that obtained by considering the XXL angular mask to assign
angular coordinates (R.A.-Dec) to the random objects. The $1\sigma$
$\Omega_{\rm M}-b_{eff}$ contours are shown in
Fig. \ref{fig:test5}. We have in this case $\Omega_{\rm
  M}=0.26_{-0.04}^{+0.05}$ and $b_{eff}=2.78_{-0.18}^{+0.17}$. The
difference with the reference case is thus within the estimated
uncertainties.

\subsection{Modelling assumptions}
\label{subsec:fitting}

To compute the effective bias of the XXL cluster sample, we assumed
the \citet{tinker2010} model in Eq. \ref{eq:bias_eff} (see Section
\ref{sec:cosmological}). To check the impact of this assumption, we
repeated our statistical analysis assuming the
\citet{sheth_mo_tormen2001} bias, converting the XXL masses to
$M_{200}$. The result, shown in Fig.\ref{fig:test5}, is fully
consistent with that obtained using the \citet{tinker2010} model,
demonstrating that our conclusions are robust with respect to the bias
model adopted. Specifically, we obtain in this case $\Omega_{\rm
  M}=0.25_{-0.04}^{+0.06}$ and $b_{eff}=2.66_{-0.15}^{+0.14}$.

The reference fitting analysis has been performed in the scale range
$10<r[$\Mpch$]<40$. We show in Fig.\ref{fig:test5} the $1\sigma$
$\Omega_{\rm M}-b_{eff}$ confidence contours obtained by enlarging the
fitting range to $1<r[$\Mpch$]<50$. The best-fit values are
$\Omega_{\rm M}=0.29_{-0.04}^{+0.05}$ and
$b_{eff}=2.67_{-0.17}^{+0.16}$, consistent with the reference case.


\subsection{Mass uncertainties}
\label{subsec:masses}

As described in Section \ref{sec:cosmological}, we used the XXL
cluster masses to estimate the effective bias of the sample
(Eq. \ref{eq:bias_eff}). The mass measurements depend on the
cosmological model. However, given the current clustering
uncertainties, this dependence can be safely neglected. To estimate
the impact of this assumption on the error budget, we repeated our
analysis converting the masses from the assumed cosmology to the test
values at each MCMC step, using Eq. C4 in \citet{sereno2015b}. The
difference in the $\Omega_M$ best-fit value with respect to the
reference case is less than $1\%$ of the estimated error.

The given mass uncertainties considered in our computations include
only the statistical errors due to the count rate. To check the impact
of this assumption, we performed our statistical analysis by
progressively increasing the value of the statistical mass errors. We
find that the best-fit value of $\Omega_M$ shifts systematically to
higher values as the mass errors increase, though the impact is
marginal. In fact, even doubling the mass errors, the effect is below
$1-\sigma$, as shown in Fig. \ref{fig:test5}. In this case, we obtain
$\Omega_{\rm M}=0.29_{-0.04}^{+0.06}$ and
$b_{eff}=2.78_{-0.18}^{+0.16}$.

While the uncertainty on the statistical errors is thus not an issue,
systematic errors on cluster masses \citep[see
  e.g.][\citetalias{eckert2016}]{eckert2016}, if present, can more
severely impact our cosmological constraints. Specifically, we find a
systematic shift to lower values of $\Omega_M$ for a systematic error
that increases the masses. As an illustrative case, in
Fig. \ref{fig:test5} we show how our cosmological constraints change
if we assume that all the XXL masses are overestimated by $20\%$. In
this case, we obtain $\Omega_{\rm M}=0.23_{-0.03}^{+0.05}$ and
$b_{eff}=2.48_{-0.16}^{+0.15}$. This highlights the importance of
having a good knowledge of any systematics possibly affecting the
cluster mass measurements. Nevertheless, even in this quite extreme
case, the effect on $\Omega_M$ is within $1-\sigma$.


\bibliography{bib}

\end{document}